\documentstyle[aps,multicol,epsf]{revtex}
\epsfverbosetrue

\title{Non-Hermitian Delocalization and Eigenfunctions}
\author{Naomichi Hatano}
\address{Theoretical Division, MS-B262, Los Alamos National Laboratory,
Los Alamos, New Mexico 87545}
\author{David R. Nelson}
\address{Lyman Laboratory of Physics, Harvard University,
Cambridge, Massachusetts 02138}
\date{May 15, 1998}

\newcounter{figurenumber}
\newcounter{figurenumberstore}
\renewenvironment{figure}%
  {\begin{list}{}{
       \setlength{\topsep}{0in}
       \setlength{\leftmargin}{0in}
       \setlength{\rightmargin}{0in}
       \setlength{\itemindent}{\parindent}
       \setlength{\labelsep}{0in}
       \usecounter{figurenumber}
       \setcounter{figurenumber}{\value{figurenumberstore}}
       \addtocounter{figurenumber}{-1}
       \stepcounter{figurenumberstore}
     }
  }{\end{list}}
\renewcommand{\caption}[1]{\refstepcounter{figurenumber}
\item FIG.~\arabic{figurenumber}.\ #1}

\begin{document}

\maketitle

\begin{abstract}
Recent literature on delocalization in non-Hermitian systems has 
stressed criteria based on sensitivity of eigenvalues to boundary 
conditions and the existence of a non-zero current.
We emphasize here that delocalization {\em also} shows up clearly in 
eigenfunctions, provided one studies the product of left- and 
right-eigenfunctions, as required on physical grounds, and {\em not} 
simply the squared modulii of the eigenfunctions themselves.
We also discuss the right- and left-eigenfunctions of the ground state 
in the delocalized regime and suggest that the behavior of these functions, 
when considered separately, may be viewed as ``intermediate'' 
between localized and delocalized.
\end{abstract}


\begin{multicols}{2}

\section{Introduction}
\label{sec1}

A delocalization phenomenon in a particularly simple class of 
non-Hermitian random system has attracted considerable attention
recently~\cite{Hatano96,Hatano97,Chen96,%
Shnerb97,Efetov97,Feinberg97a,Janik97c,Brouwer97,%
Feinberg97c,Goldsheid97,Brezin97,Nelson97,Feinberg97d,Janik97e,Zee97,%
Hatano98,Shnerb97b,Mudry97,Shnerb98}.
Among the more recent work is a report by Silvestrov~\cite{Silvestrov98},
based on an analysis of eigenfunctions,
which claims that the phenomenon studied was not actually 
delocalization, but ``localization of a very unusual kind.''
Although Silvestrov subsequently revised his views, he still 
maintains that ``the transition from real to complex spectra in 1d 
disordered systems with (an) imaginary vector potential is not a 
delocalization transition''~\cite{Silvestrov98-2}.
In this paper, we review some basic facts of the non-Hermitian 
delocalization (Sec.~\ref{sec2}) and then
take issue with Silvestrov's interpretation.
We stress in Sec.~\ref{sec3} that 
the criteria for delocalization used in Refs.~\cite{Hatano96,Hatano97} 
are entirely consistent with a conventional one based on 
eigenfunctions, provided one studies the correct physical quantity, 
namely the product of the left- and 
right-eigenfunctions associated with a given state.
In Sec.~\ref{sec4}, we comment 
on the interesting results of Silvestrov for left- 
and right-eigenfunctions considered separately for large asymmetry 
parameter.
We show for the 
ground state that the results are related to earlier results 
obtained for charge density waves~\cite{Chen96} and population 
biology~\cite{Nelson97}.
From this viewpoint, we argue that the behavior of
the left- and the right-eigenfunctions is ``intermediate'' 
between localized and delocalized behavior.

\section{Non-Hermitian delocalization: Eigenvalues and current}
\label{sec2}

Let us first review some basic facts about non-Hermitian 
delocalization.
A typical example of the systems in question is the 
one-particle Hamiltonian
\begin{equation}
\label{eq:ham1D}
{\cal H}=\frac{(p+ig)^{2}}{2m}+V(x),
\end{equation}
where $p$ is the momentum operator $-i\hbar d/dx$,
$g$ is a non-Hermitian field constant in time and space, and
$V(x)$ is a random potential.
Its lattice version is given by the matrix
\begin{equation}
\label{eq:ham-mat}
{\cal H}_{xx'}=-\frac{t}{2}
\left(e^{\bar{g}}\delta_{x,x'+1}+e^{-\bar{g}}\delta_{x,x'-1}\right)
+V_{x}\delta_{x,x'},
\end{equation}
where $x$ and $x'$ here are site indices, $V_{x}$ is a random 
potential, and $\bar{g}=ga/\hbar$ with $a$ denoting the lattice spacing.
For simplicity, we focus on the one-dimensional case throughout this 
paper.
Periodic boundary conditions are imposed except 
where stated otherwise.
The above Hamiltonian reduces to the Anderson localization problem for 
$g=0$;
in this case,
it is widely believed that all eigenfunctions are localized in one 
and two dimensions.

We showed~\cite{Hatano96,Hatano97} that eigenvalues become complex
pair by pair once $g$ is increased beyond a threshold value $g=g_{c1}$
and that the states with complex eigenvalues are delocalized.
To show the delocalization, we presented two pieces of evidence.
First, we numerically showed that the states with complex eigenvalues 
carry a current.
The current carried by the $n$th eigenstate is defined by
$j_{n}=\partial \varepsilon_{n} (g)/\partial (ig)$, where 
$\varepsilon_{n}$ is the eigenvalue.
This is the standard definition of the current, because $g$ in 
Eqs.~(\ref{eq:ham1D}) and~(\ref{eq:ham-mat}) plays a role of imaginary 
vector potential.
The current was clearly nonzero for 
states in the bubble of complex eigenvalues in the band center (see 
Fig.~13 of Ref.~\cite{Hatano97} and Fig.~\ref{fig:spectrum}~(b) below),
indicating the delocalization of the states.

As a second indication of delocalization, 
we showed that the delocalized states have complex 
eigenvalues for systems with periodic boundaries, but that
all eigenvalues remain real when the same system has open boundary 
conditions.
This sensitivity to boundary conditions is another indication that the 
corresponding wave functions are delocalized.
We confirmed these two signatures of delocalization in a sufficiently 
strong imaginary vector
potential with numerical work and analytic 
calculations on localized impurities.

\begin{figure}
\epsfxsize=3.375in
\epsfbox{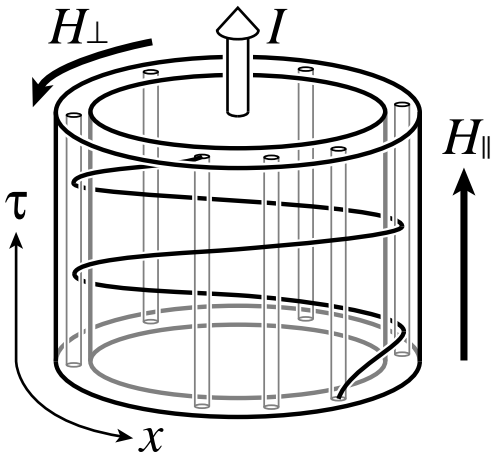}
\caption{Vortex-line system characterized by
a one-dimensional periodic non-Hermitian transfer matrix.
A magnetic field  $H_{\parallel}$ forces a flux line into a 
cylindrical shell of type-II superconductor with columnar defects.
The current threading the cylinder generates the magnetic-field
component $H_{\perp}$, which tries to tilt the flux line.
When this flux-line system is mapped onto a ring of the non-Hermitian 
system, $H_{\perp}$ becomes proportional to the non-Hermitian field $g$.
Below a certain strength of $H_{\perp}$ (or $g$),
the flux line is pinned by a columnar defect and forced to run parallel to 
the defects (the transverse Meissner effect~\protect\cite{Nelson93}), 
except for its slight deflection from the pinning 
center near the top and the bottom of the 
cylinder~\protect\cite{Hatano97}.
For large enough  $H_{\perp}$, however, the flux line is depinned and wraps 
around the cylinder as is shown here as a helix; 
This gives rise to a nonzero current that circulates
around the ring of the corresponding non-Hermitian system.}
\label{fig:cylinder}
\end{figure}

This delocalization phenomenon is equivalent to 
flux-line depinning in type-II superconductors with extended defects.
Suppose that a superconductor has columnar defects randomly located 
but mutually parallel and that an external magnetic field forces a flux 
line into the superconductor.
The flux line tends to be pinned by a columnar defect (or a 
collection of them) when the external field is parallel to the defects.
When the field is tilted away from the axis of the defects,
we expect flux-line depinning at a certain tilt angle
(Fig.~\ref{fig:cylinder}); 
see Refs.~\cite{Jiang94,Safar96,Obaidat97} for experiments.
\vskip 0.3in
\vskip 0.3in

The physics of vortex matter can be mapped onto quantum systems with 
one less dimension
by the inverse of the Feynman path integral mapping~\cite{Nelson93};
that is, we regard the Boltzmann weight of the flux line as an
exponentiated action of the world line of a quantum particle and make 
the identification $\hbar\longleftrightarrow T$, where $T$ is the 
temperature of the vortex system.
This procedure gives Hamiltonians of the above type.
The component of the external magnetic field perpendicular to the 
columns is proportional to the non-Hermitian field $g$.
Depinning of the flux line by tilting the field beyond a certain 
strength of $g$ leads to a nonzero current in the corresponding 
quantum state~\cite{Hatano97}.

\section{Non-Hermitian delocalization: Wave functions}
\label{sec3}

Delocalization of the eigenfunctions themselves was not studied 
directly in Refs.~\cite{Hatano96,Hatano97}.
The main purpose of the present paper is to address this issue.

Silvestrov computed the right-eigenfunctions associated with the 
model~(\ref{eq:ham-mat}) for a 300-site lattice in the region of 
complex eigenvalues and found that their squared modulii have a 
sparse set of well-separated peaks, quite different from a 
conventional delocalized state~\cite{Silvestrov98}.
However, as shown in Ref.~\cite{Hatano97}, and acknowledged later by 
Silvestrov~\cite{Silvestrov98-2}, it is the product of left- and 
right-eigenvectors which determines the probability distribution for a 
tilted vortex line interacting with columnar defects deep within the 
sample.
It is this product which clearly delocalizes in the conventional 
sense when the eigenvalues become complex.
In the hope of avoiding further confusion, we first summarize in this
section the basic relation between left- and right- eigenvectors. 
We then illustrate the delocalization of their product
with numerical examples from our own extensive 1000-site-lattice 
computations.

\subsection{Left- and right-eigenfunctions}

We work for concreteness with the continuum 
Hamiltonian~(\ref{eq:ham1D}), but the results also apply to lattice 
non-Hermitian models like~(\ref{eq:ham-mat}).
Suppose we have computed a set $\{\phi_{n}^{R}(x;g)\}$ of 
right-eigenfunctions of ${\cal H}(g)$ which satisfy
\begin{equation}\label{eq:3}
  {\cal H}(g) \left| n;g \right\rangle
  =\varepsilon_{n}(g) \left| n;g \right\rangle,
\end{equation}
where we adopt the Dirac bra-ket notation,
\begin{equation}\label{eq:4}
  \phi_{n}^{R}(x;g) \longrightarrow  \left| n;g \right\rangle.
\end{equation}
Although left-eigenvectors need not be simply related to 
right-eigenvectors in general, there is a particularly simple relation 
for the Hamiltonian~(\ref{eq:ham1D}), which arises due to the 
symmetry~\cite{Hatano97}
\begin{equation}\label{eq:5}
  {\cal H}^{\dag}(g) = {\cal H}(-g),
\end{equation}
where $\dag$ denotes the usual Hermitian conjugate.
Indeed, as shown below, left-eigenvectors can be obtained from 
right-eigenvectors by complex conjugation and letting $g\longrightarrow-g$,
\begin{equation}\label{eq:6}
  \phi_{n}^{L}(x;g) = \phi_{n}^{R}(x;-g)^{\ast},
\end{equation}
or in Dirac notation,
\begin{equation}\label{eq:7}
  \phi_{n}^{L}(x;g) \longrightarrow \left\langle n;g \right| 
  \equiv \left| n;-g \right\rangle^{\dag},
\end{equation}
where $\dag$ again denotes conventional Hermitian conjugation.
Our convention that the left-eigenvector $\left\langle n;g \right|$ is defined 
to be the Hermitian conjugate of $\left| n;-g \right\rangle$, {\em not} 
of $\left| n;g \right\rangle$, allows manipulations which parallel closely
those of conventional quantum mechanics.
To see that $\left\langle n;g \right|$ is in fact a 
left-eigenfunction, we calculate
\begin{eqnarray}\label{eq:8}
    \left\langle n;g \right| {\cal H}(g) & = & 
    \left( {\cal H}(g)^{\dag} \left| n;-g \right\rangle \right)^{\dag}
  \nonumber
  \\
     & = & 
    \left( {\cal H}(-g) \left| n;-g \right\rangle \right)^{\dag}
  \nonumber
  \\
	 & = & \left\langle n;g \right| \varepsilon_{n}(-g)^{\ast}.
\end{eqnarray}
Evidently, $\left\langle n;g \right|$ will indeed be a 
left-eigenfunction with the {\em same} eigenvalue as 
$\left| n;g \right\rangle$, provided
\begin{equation}\label{eq:9}
  \varepsilon_{n}(-g)^{\ast} = \varepsilon_{n}(g)
\end{equation}

To prove Eq.~(\ref{eq:9}), we let ${\cal H}(g)$ act to the right and 
the left in the matrix element
\begin{equation}\label{eq:10}
  \left\langle m;g \right|\, {\cal H}(g) \,\left| n;g \right\rangle
\end{equation}
and obtain
\begin{equation}\label{eq:11}
  \left[\varepsilon_{m}(-g)^{\ast}-\varepsilon_{n}(g)\right]
  \left\langle m;g \,|\, n;g \right\rangle=0.
\end{equation}
Equation~(\ref{eq:9}) follows by setting $m=n$, provided 
$\left\langle n;g \,|\, n;g \right\rangle \neq 0$.
More generally, Eq.~(\ref{eq:11}) can be used to show that, with proper 
normalization, the right- and left-eigenfunctions form a biorthogonal 
set,
\begin{equation}\label{eq:12}
  \left\langle m;g \,|\, n;g \right\rangle = \delta_{m,n}.
\end{equation}
This set has the usual completeness relation
\begin{equation}\label{eq:13}
  \sum_{n}\left| n;g \right\rangle \left\langle n;g \right| = 1.
\end{equation}
Equation~(\ref{eq:9}) reduces to the usual Hermitian constraint of real 
eigenvalues when $g=0$.

Once the eigenvectors are properly normalized,
the imaginary-time particle propagator is given by
\begin{equation}
G(\tau)=\sum_{n}\left| n;g \right\rangle \left\langle n;g \right|
e^{-\varepsilon_{n}(g) \tau/\hbar},
\end{equation}
or, in the coordinate representation,
\begin{equation}
G(x,x';\tau)=\sum_{n}\phi_{n}^{R}(x)\phi_{n}^{L}(x')
e^{-\varepsilon_{n}(g)\tau/\hbar}.
\end{equation}
The density distribution of a particle in the ground state (which 
dominates as $\tau\to\infty$) is
hence the product of left- and right-eigenfunctions,
$\phi_{\rm gs}^{L}(x)\phi_{\rm gs}^{R}(x)$.
As was shown in Ref.~\cite{Hatano97}, this product gives the probability 
distribution of a flux line far from the sample 
boundaries in the imaginary-time direction (the top and the bottom 
edges of the cylinder in Fig.~\ref{fig:cylinder}).
The square modulii $|\phi_{n}^{R}(x)|^{2}$ and $|\phi_{n}^{L}(x)|^{2}$ 
are {\em irrelevant} for the bulk properties.

\subsection{Delocalization of $\phi^{L}\phi^{R}$}

\begin{figure}
\epsfxsize=3.375in
\epsfbox{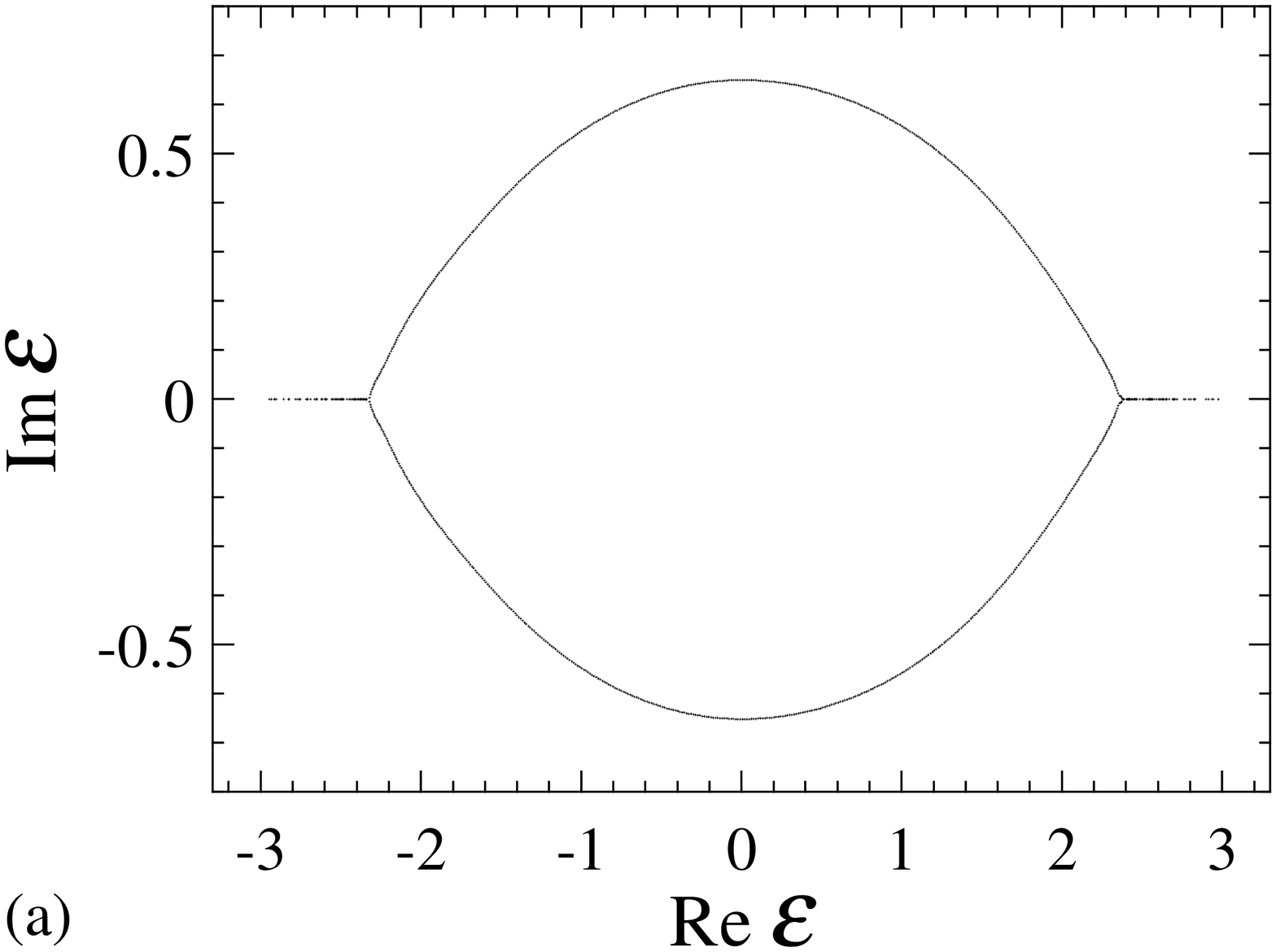}
\epsfxsize=3.375in
\epsfbox{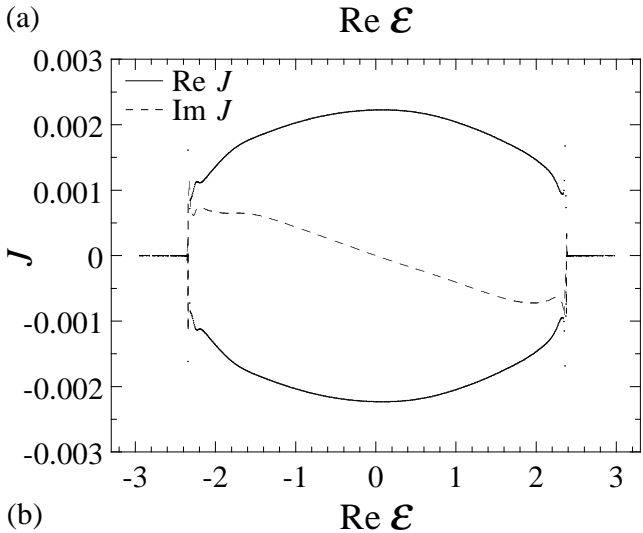}
\caption{The complex energy spectrum and the current distribution 
of the Hamiltonian~(\protect\ref{eq:ham-mat}) on a 1000-site lattice.
Each eigenvalue is marked by a tiny cross in (a).
Each pair of complex conjugate eigenvalues in (a) has the current shown
in (b), with
the reals part of the opposite sign (tiny crosses) and 
the identical imaginary parts (the dashed line).
The parameters in Eq.~(\protect\ref{eq:ham-mat})
are set to $t=2$ and $\bar{g}=0.4$ with 
each $V_{x}$ chosen randomly from the range $[-1.5,1.5]$.}
\label{fig:spectrum}
\end{figure}

We now illustrate the different behaviors of $|\phi_{n}^{R}(x)|^{2}$,
$|\phi_{n}^{L}(x)|^{2}$ and $\phi_{n}^{L}(x)\phi_{n}^{R}(x)$ 
with numerical results,
emphasizing that the product $\phi_{n}^{L}\phi_{n}^{R}$ is clearly 
delocalized in the conventional sense when eigenvalues become complex.
We consider a particular realization of the random
Hamiltonian~(\ref{eq:ham-mat}) on a 1000-site lattice.
The parameters are set to $t=2$ and $\bar{g}=0.4$ with 
the value of $V_{x}$ at each site chosen randomly from the range 
$[-1.5,1.5]$.
These values are the same as used in 
Refs.~\cite{Silvestrov98,Silvestrov98-2} except 
that the system size is greater in our calculation.
(Note that the definition of $t$ differs by factor two.)
The energy spectrum is shown in Fig.~\ref{fig:spectrum}~(a).
The states between the two mobility edges 
$\varepsilon_{c}\simeq\pm 2.34$
have complex eigenvalues (and hence we would argue are delocalized), 
while the other states are localized.
Every delocalized states carries a complex current as is shown in 
Fig.~\ref{fig:spectrum}~(b).
The imaginary part of the current determines the tilt angle of
a flux line~\cite{Hatano96,Hatano97}.

\begin{figure}
\epsfxsize=3.375in
\epsfbox{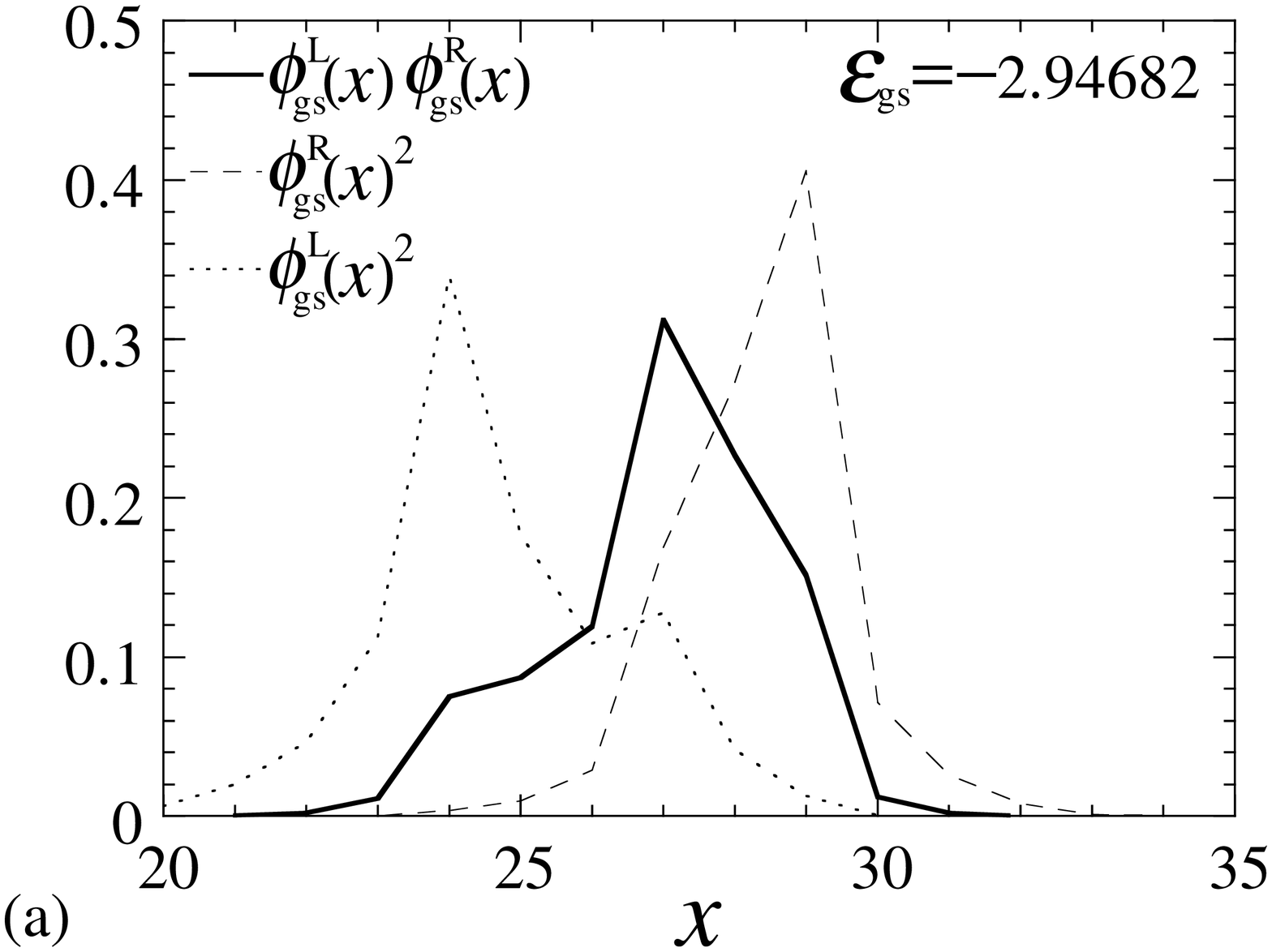}
\epsfxsize=3.375in
\epsfbox{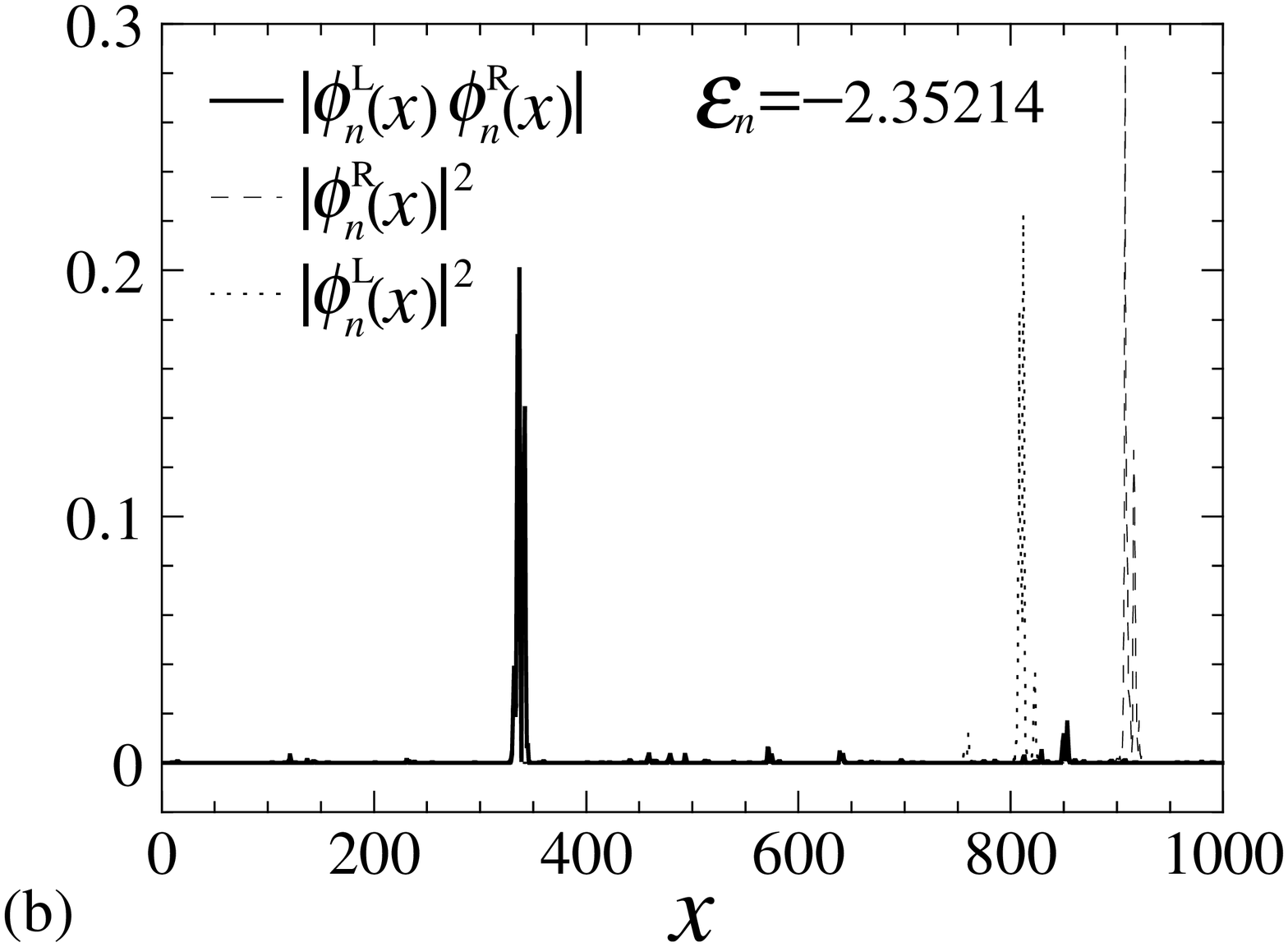}
\epsfxsize=3.375in
\epsfbox{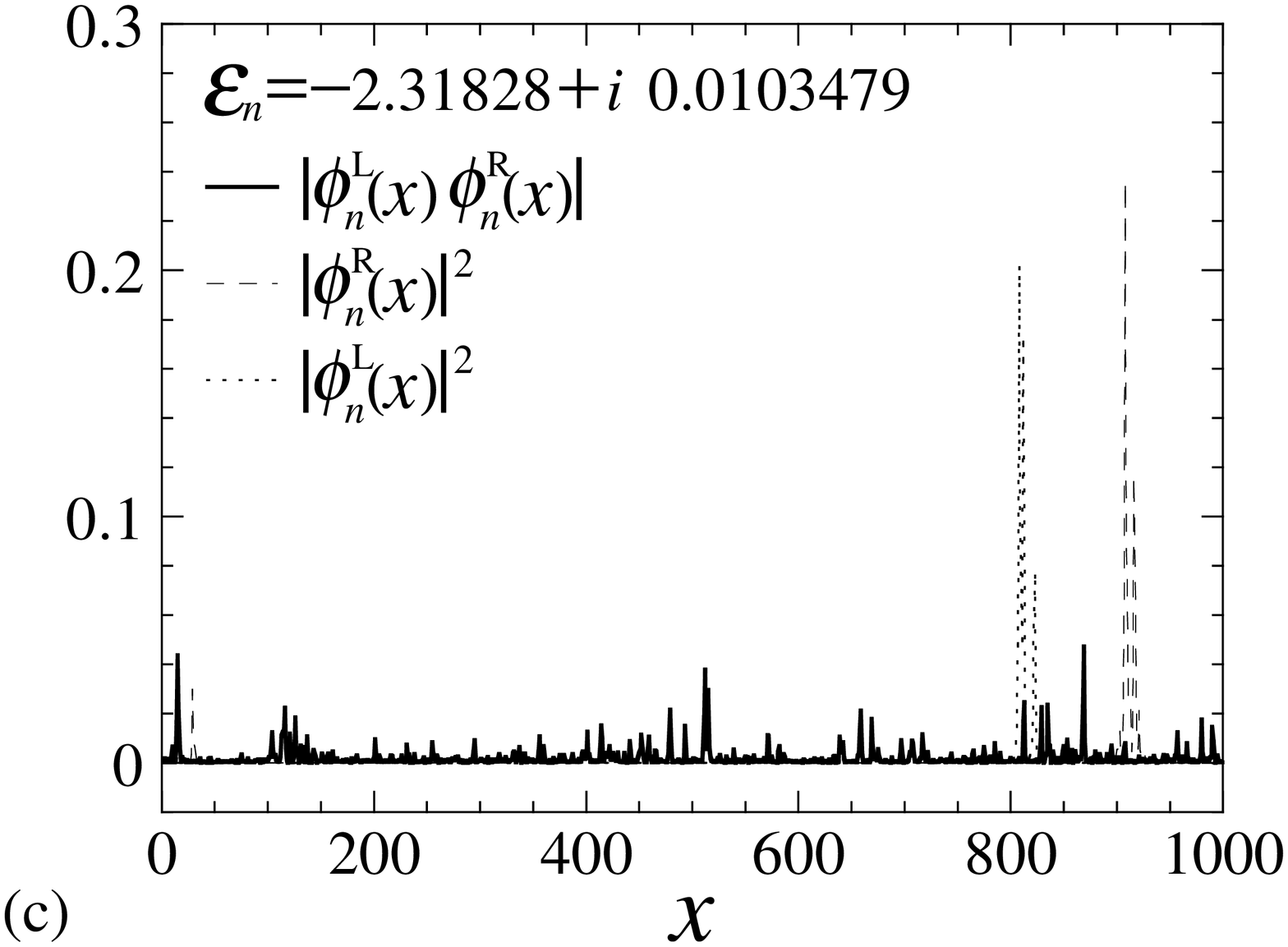}
\caption{The functions 
$|\phi^{L}(x)\phi^{R}(x)|$ (thick solid lines), 
$|\phi^{R}(x)|^{2}$ (dashed lines) and
$|\phi^{L}(x)|^{2}$ (dotted lines)
for the following cases: 
(a) the ground state
($\varepsilon=-2.94682$);
(b) an eigenstate just below the lower mobility edge
(the 72nd state with $\varepsilon=-2.35214$);
(c) an eigenstate just above the lower mobility edge
(the 80th state with $\varepsilon=-2.31828+i\,0.0103479$).
The system is the same as the one used in Fig.~\protect\ref{fig:spectrum}.
The serial number of each state represents the ascending order of the 
real part of the eigenvalue.}
\label{fig:wf}
\end{figure}

Figure~\ref{fig:wf}~(a) shows the functions 
$\phi_{n}^{L}(x)\phi_{n}^{R}(x)$, 
$|\phi_{n}^{R}(x)|^{2}$ and
$|\phi_{n}^{L}(x)|^{2}$
for the (localized) ground state.
All quantities are 
normalized so that the summation over $x$ yields unity.
We stress, however, that the normalization only makes physical sense for
$\phi_{n}^{L}(x)\phi_{n}^{R}(x)$.
Everywhere in the regime of localized states,
\begin{eqnarray}\label{eq:gauge}
  \phi_{n}^{R}(x;g) & \propto & e^{gx/\hbar}\phi_{n}(x;0)
  \nonumber
  \\
  \mbox{and}\qquad
  \phi_{n}^{L}(x;g) & \propto & e^{-gx/\hbar}\left(\phi_{n}(x;0)\right)^{\ast}
\end{eqnarray}
for large enough systems,
where $\phi_{n}(x;0)$ is the wave function of the Hamiltonian 
with $g=0$.
Note that Eq.~(\ref{eq:6}) is obeyed.
(The specific $g$-dependence in Eq.~(\ref{eq:gauge}) {\em only} holds 
for localized states, 
for which we can always choose $\phi_{n}(x;0)$ to be real.)
Hence, the product $\phi_{n}^{L}(x;g)\phi_{n}^{R}(x;g)=|\phi_{n}(x;0)|^{2}$ 
does not depend on $g$ until the state is delocalized for large 
enough $g$.
This is a mathematical expression of the transverse Meissner effect, or
the rigidity of the pinned flux line against the tilt of the applied 
magnetic field; see Refs.~\cite{Hatano97,Nelson93} for details.

Figure~\ref{fig:wf}~(b) shows the functions 
$|\phi_{n}^{L}(x)\phi_{n}^{R}(x)|$, 
$|\phi_{n}^{R}(x)|^{2}$ and
$|\phi_{n}^{L}(x)|^{2}$
for a state slightly below the lower mobility edge, 
while Fig.~\ref{fig:wf}~(c) shows those for
a state slightly above the edge.
(We only plot the amplitude of the function $\phi_{n}^{L}(x)\phi_{n}^{R}(x)$;
the phase oscillates rapidly for delocalized states away from the 
band edges.)
The function $|\phi_{n}^{L}(x)\phi_{n}^{R}(x)|$ changes dramatically across 
the mobility edge, while the changes in $|\phi_{n}^{R}(x)|^{2}$ and
$|\phi_{n}^{L}(x)|^{2}$ are less noticeable.
\nopagebreak

\begin{figure}
\epsfxsize=3.375in
\epsfbox{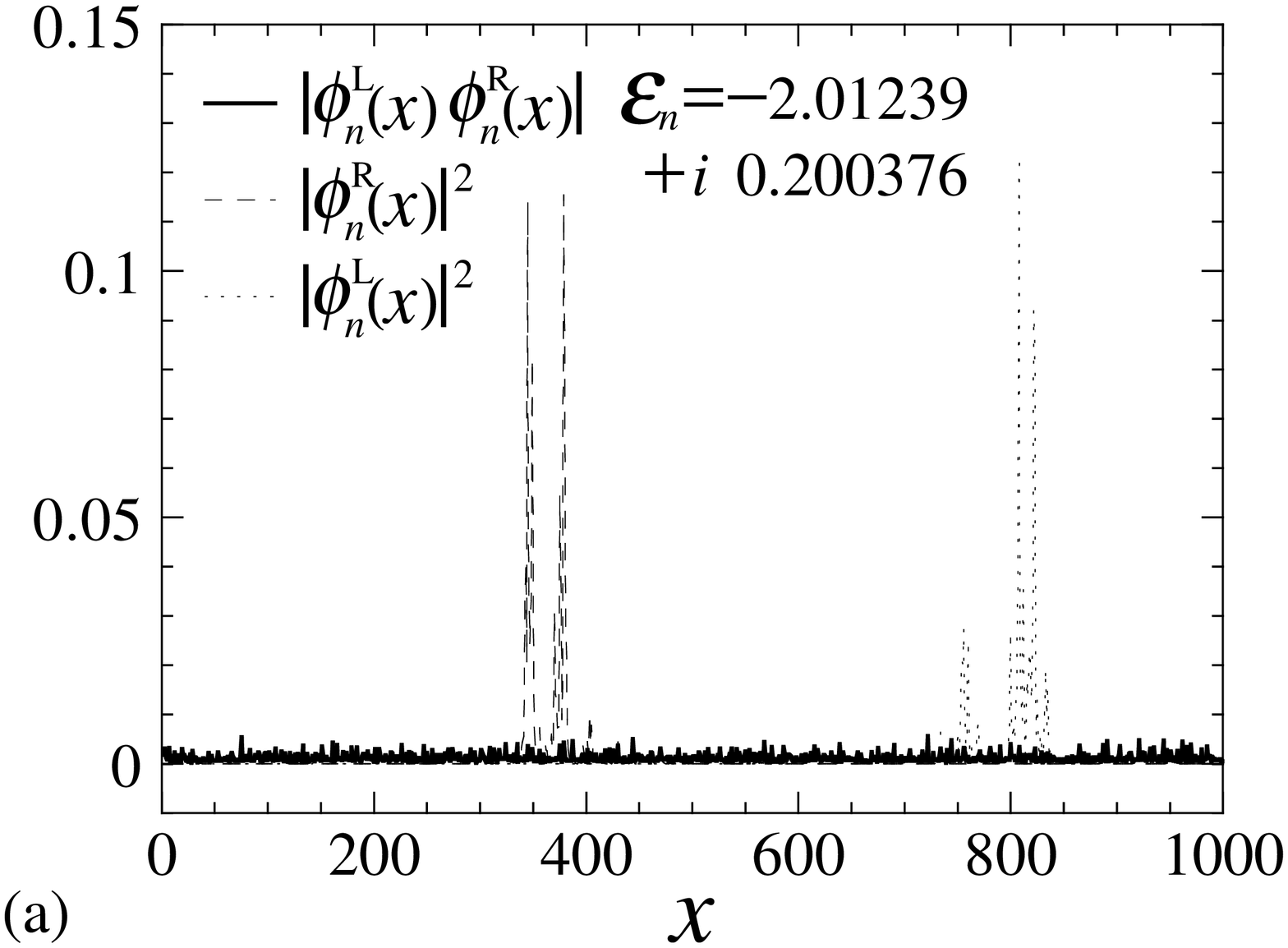}
\epsfxsize=3.375in
\epsfbox{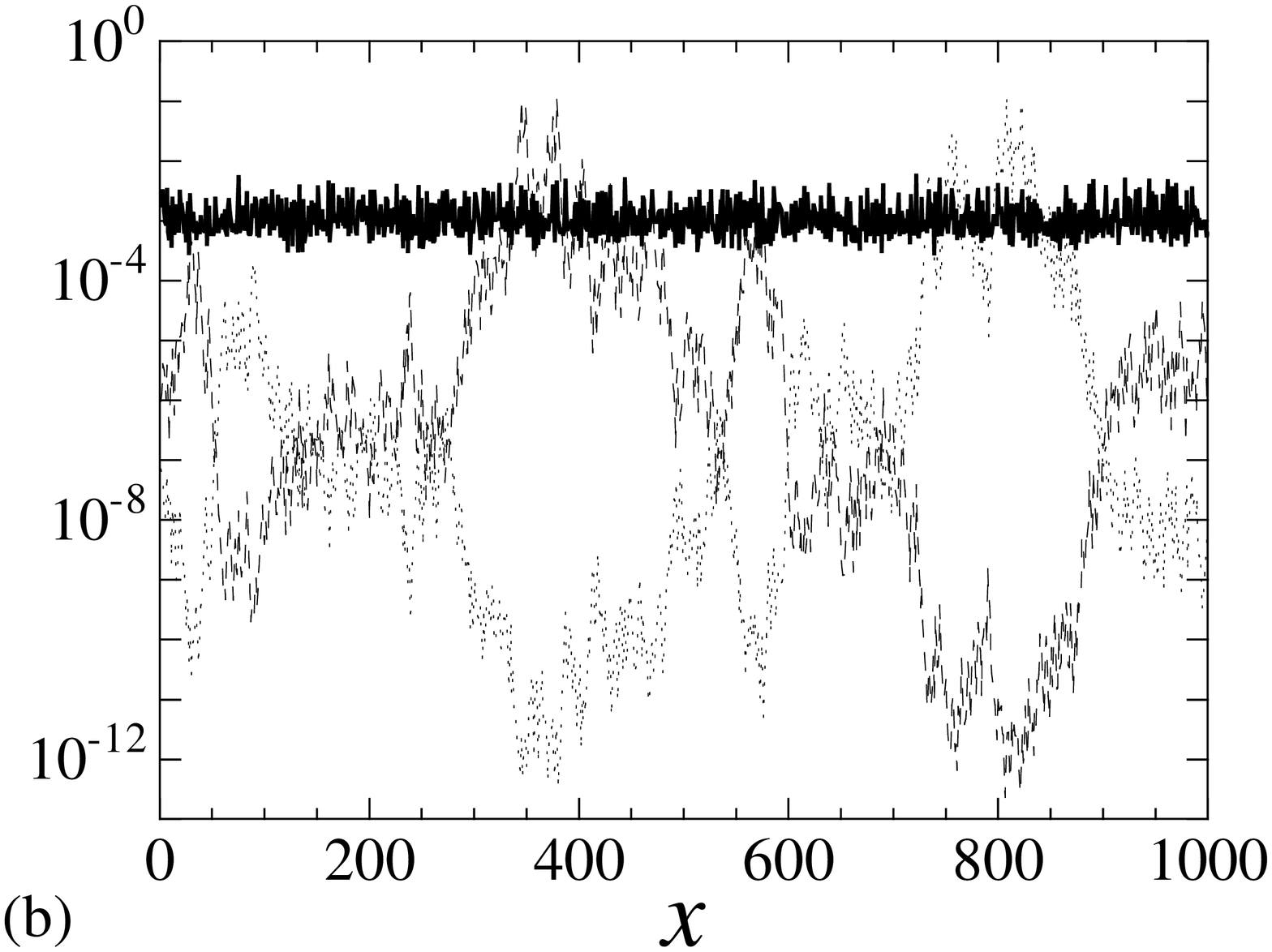}
\caption{The functions 
$|\phi^{L}(x)\phi^{R}(x)|$ (thick solid lines), 
$|\phi^{R}(x)|^{2}$ (dashed lines) and
$|\phi^{L}(x)|^{2}$ (dotted lines)
for the an eigenstate in the delocalized regime
(the 166th state with $\varepsilon=-2.01239+i\,0.200376$);
(a) a linear plot and (b) a semi-logarithmic plot.
The system is the same as the one used in Fig.~\protect\ref{fig:spectrum}.}
\label{fig:wf-2}
\end{figure}

\nopagebreak
The delocalized nature of $\phi_{n}^{L}(x)\phi_{n}^{R}(x)$ appears even more 
dramatically deep inside the bubble of complex eigenvalues.
Figure~\ref{fig:wf-2}~(a) shows the same functions 
for an eigenstate with the eigenvalue 
$\varepsilon_{n}=-2.01239+i\,0.200376$.
This is the 166th state, which roughly corresponds to the 
50th state of the 300-site system studied by 
Silvestrov~\cite{Silvestrov98,Silvestrov98-2}.
In Fig.~\ref{fig:wf-2}~(a), the function $|\phi_{n}^{L}(x)\phi_{n}^{R}(x)|$ 
is extended and approximately constant, while 
$|\phi_{n}^{R}(x)|^{2}$ and $|\phi_{n}^{L}(x)|^{2}$ exhibit a sparse set of 
well-separated maxima.
Following Silvestrov, we plot the logarithm of these functions in 
Fig.~\ref{fig:wf-2}~(b).
The {\em product} of $\phi^{R}(x)$ and $\phi^{L}(x)$ is 
remarkably constant and is extended in conventional sense.
On the other hand, the ragged wandering nature of 
$\ln |\phi_{n}^{R}(x)|$ and 
$\ln |\phi_{n}^{L}(x)|$ is consistent with the 
conjecture~\cite{Silvestrov98,Silvestrov98-2} that these functions 
behave like random walks as a function of $x$;
this is the subject of the next section.

\vskip 0.2in

\section{Random-walk behavior of eigenfunctions}
\label{sec4}

In this section, we turn our attention to the left- and 
right-eigenfunctions considered separately for large $g$.
One of the interesting results of 
Silvestrov~\cite{Silvestrov98,Silvestrov98-2} 
is a random-walk-like behavior hidden in the logarithms of the modulii
of these eigenfunctions.
We first illustrate the random-walk behavior with our more 
extensive numerical results and then show that, at least for the ground 
state, Silvestrov's observation is related to earlier results 
obtained for charge density waves~\cite{Chen96} and population 
biology~\cite{Nelson97}.
As a concluding remark, we argue that the behavior of
sample-to-sample fluctuations of the left and right ground-state
eigenfunctions considered separately for large $g$
is ``intermediate'' between that expected for localized and 
delocalized states.

\subsection{Vortex-line distribution at boundaries}

What information is contained in the functions $\phi_{n}^{R}(x)$ and 
$\phi_{n}^{L}(x)$ (considered separately) for flux line systems?
For a single vortex line, only the {\em ground state} contributes in 
the limit of a very long cylinder.
As discussed in Ref.~\cite{Hatano97} and exploited in a very recent 
paper by Silvestrov~\cite{Silvestrov98b}, the (nodeless) ground state wave 
functions $\phi_{\rm gs}^{R}(x)$ and $\phi_{\rm gs}^{L}(x)$ 
({\em not} their modulii squared) are proportional to the vortex-line 
probability distribution at the boundaries where it enters and leaves the 
cylinder (see Fig.~\ref{fig:cylinder}).

There are then two cases to consider.
For small and intermediate values of $g$, the spectrum is either 
completely localized or only partially delocalized as in 
Fig.~\ref{fig:spectrum}~(a).
In this case the ground state is localized as in Fig.~\ref{fig:wf}~(a),
and hence the single vortex line is pinned close to a preferred 
columnar defect in the bulk of the 
superconductor cylinder.
The right- and left-eigenfunctions are shifted relative to their
product.
This reflects the tendency of the localized vortex line to tear 
away from the pinning center at the top and bottom of the sample when 
$g$ is nonzero; see Fig.~15~(a) of Ref.~\cite{Hatano97} for a 
demonstration.

\begin{figure}
\vskip 0.4in
\epsfxsize=3.375in
\epsfbox{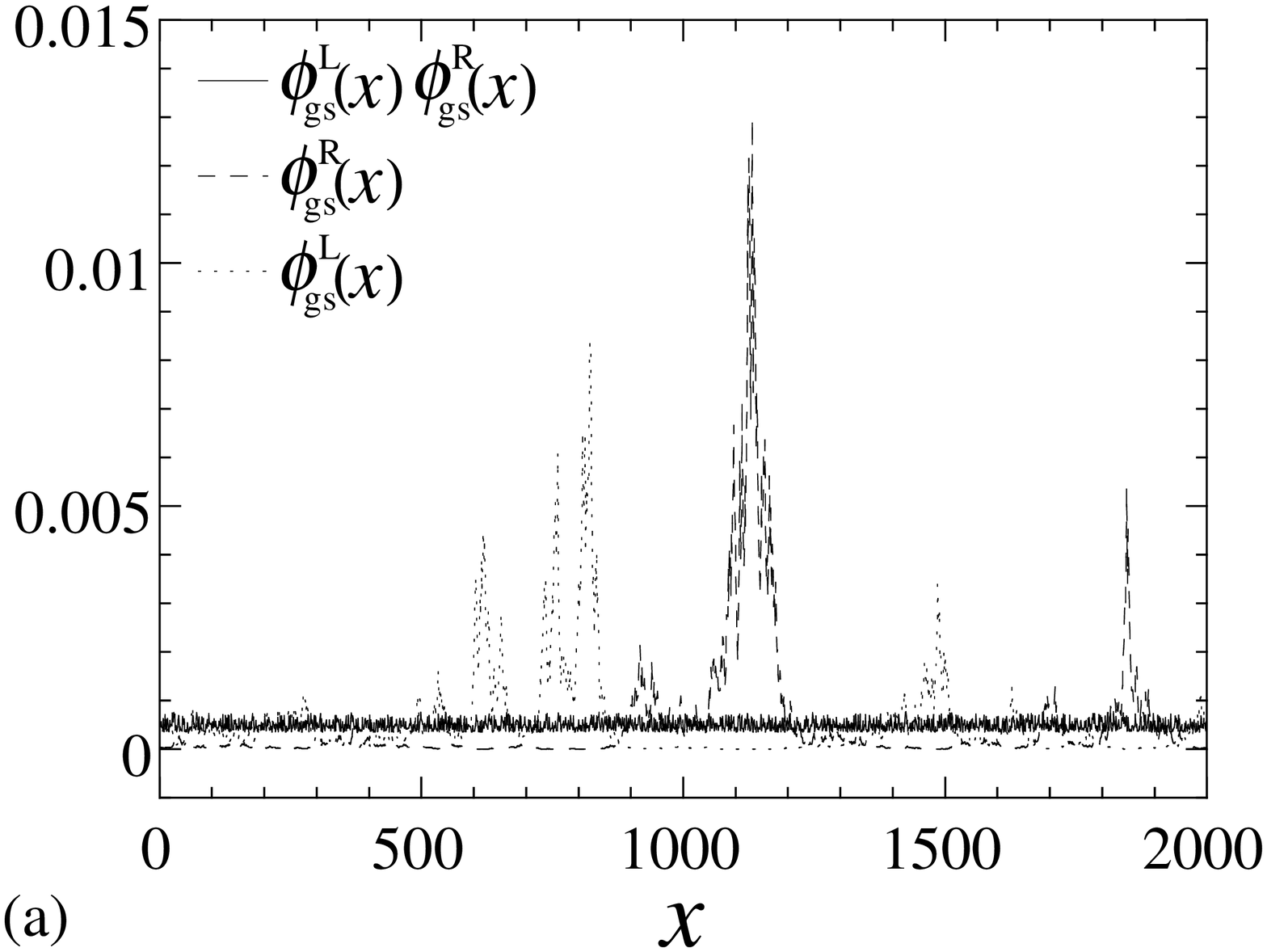}
\epsfxsize=3.375in
\epsfbox{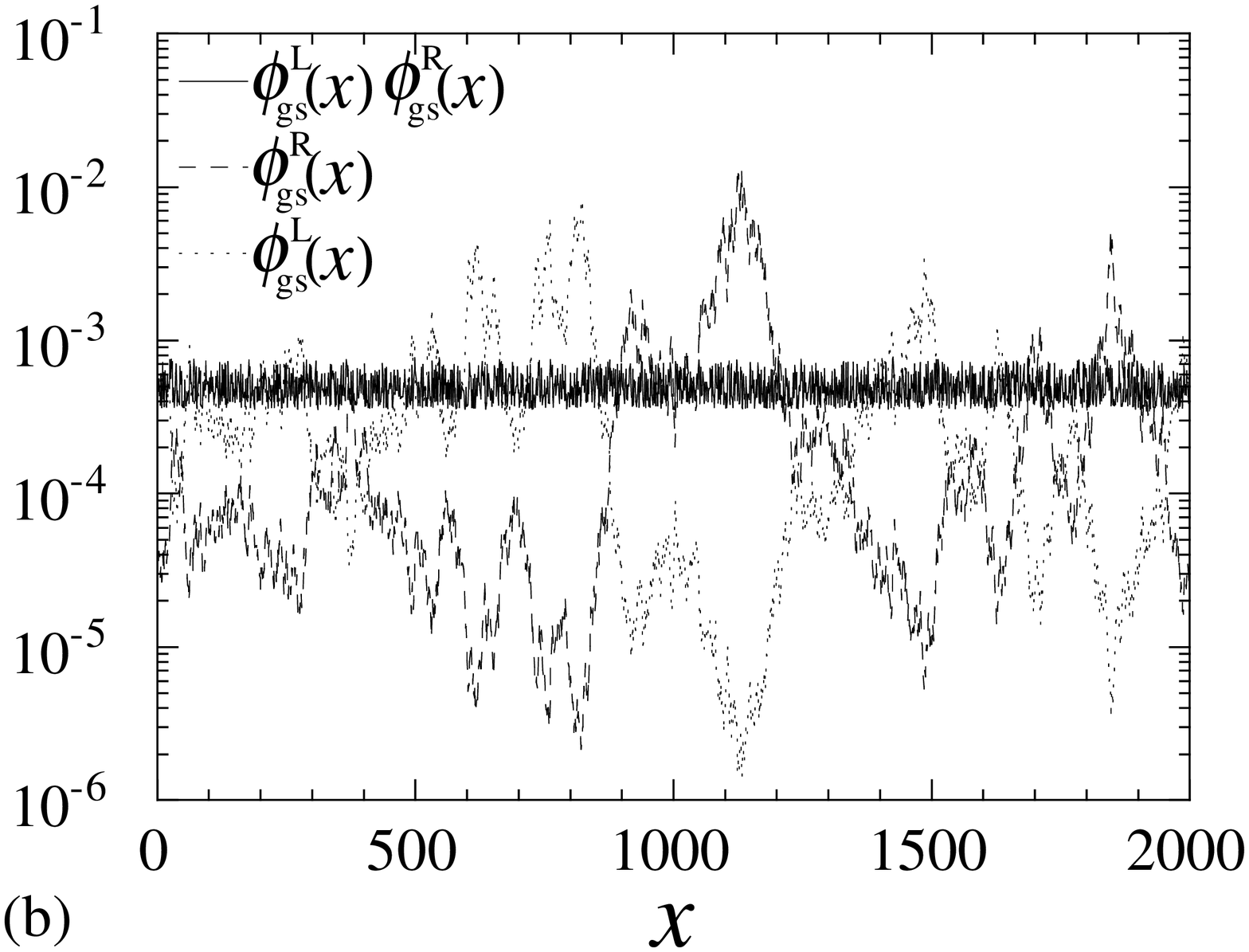}
\caption{The functions 
$\phi^{L}(x)\phi^{R}(x)$ (think solid lines), 
$\phi^{R}(x)$ (dashed lines) and
$\phi^{L}(x)$ (dotted lines)
for the delocalized ground state of a 2000-site lattice;
(a) a linear plot and (b) a semi-logarithmic plot.
All these functions are positive definite in this case.
The normalization of $\phi^{R}(x)$ and $\phi^{L}(x)$ is different from 
the one in earlier figures;
each is normalized so that its sum over $x$ (not the sum of
the squared modulus) becomes unity.
The parameter values are the same as the ones used in 
Fig.~\protect\ref{fig:spectrum} except that $\bar{g}=1.5$.}
\label{fig:wf-gs}
\end{figure}

The second more interesting case is for large $g>g_{c2}$, such that 
{\em all} states, including the ground state, are delocalized.
Using the WKB approximation, Silvestrov~\cite{Silvestrov98,Silvestrov98-2}
argued for random-walk behavior of the logarithm
of the wave functions in this case.
For concreteness, we show some of our numerical results for the 
ground state of a 2000-site lattice.
(Silvestrov~\cite{Silvestrov98-2} 
did not show numerical results for the ground state in this regime.)
Figure~\ref{fig:wf-gs}~(a) shows the ground state quantities 
$\phi_{\rm gs}^{L}(x)$, $\phi_{\rm gs}^{R}(x)$ and 
$\phi_{\rm gs}^{L}(x)\phi_{\rm gs}^{R}(x)$ for 
$g=1.5\hbar/a>g_{c2}$; the values of the
other parameters are the same as in the earlier figures.
The product $\phi_{\rm gs}^{L}(x)\phi_{\rm gs}^{R}(x)$ is 
approximately constant, while the (nodeless) eigenfunctions
$\phi_{\rm gs}^{L}(x)$ and $\phi_{\rm gs}^{R}(x)$ are quite different 
than in Fig.~\ref{fig:wf}~(a):
They exhibit {\em multiple} sharp maxima which are rather well separated.
In view of these multiple maxima, one might question whether it is 
appropriate to call such eigenfunctions 
``localized''~\cite{Silvestrov98,Silvestrov98-2}.
Figure~\ref{fig:wf-gs}~(b) shows the same ground state quantities 
in a semi-logarithmic plot.
The wandering, ragged shape of $\ln \phi_{\rm gs}^{R}(x)$ and 
$\ln \phi_{\rm gs}^{L}(x)$ indicates the random-walk behavior.

The different shapes of 
$\phi_{\rm gs}^{L}(x)\phi_{\rm gs}^{R}(x)$, $\phi_{\rm gs}^{R}(x)$ and 
$\phi_{\rm gs}^{L}(x)$ reflect the different 
optimization problems of the vortex-line configuration
in the bulk, at the top and at the bottom of the superconductor cylinder
(see Fig.~\ref{fig:cylinder}).
Since the string tension of the vortex line (the ``mass'' of
the corresponding quantum particle) is missing outside the superconductor,
the vortex line can take better advantage of the potential energy at the 
top and bottom of the sample than in the bulk;
hence 
the sharp maxima in $\phi_{\rm gs}^{L}(x)$ and $\phi_{\rm gs}^{R}(x)$.
The multiple maxima indicate that the depinned vortex line
can enter and exit the superconductor at variety of preferred locations.

The optimization problems are also different at the top and the 
bottom of the cylinder in Fig.~\ref{fig:cylinder}.
When the vortex line enters the sample from below, it is the succession of 
defects {\em counterclockwise} to the entry point which are most important.
When exiting the sample, it is the defects {\em clockwise} to the 
exit point that matter most.
Hence the peaks in $\phi_{\rm gs}^{L}(x)$ and in
$\phi_{\rm gs}^{R}(x)$ appear at very different locations.
Nevertheless, the entry and exit probability distributions are 
strongly correlated with each other, since their product is 
approximately constant.

\subsection{Renormalization group for the ground state}

In the following, let us reproduce Silvestrov's WKB 
result for delocalized states in a more controlled approximation.
For the ground state, his result is in fact a special case of 
numerical, scaling and renormalization group calculations applied 
previously to related problems in one-dimensional
charge density waves~\cite{Chen96} 
and population biology in $d$ dimensions~\cite{Nelson97}.

We start with the time-dependent Schr\"{o}dinger equation for the 
continuum Hamiltonian~(\ref{eq:ham1D}),
\begin{eqnarray}\label{eq:18}
    \hbar\frac{\partial\psi^{R}}{\partial\tau}(x,\tau)  & = & 
    - {\cal H}\psi^{R}(x,\tau)
      \nonumber
  \\
     & = & \frac{1}{2m}
     \left( \hbar\frac{\partial}{\partial x} -g \right)^{2} \psi^{R}(x,\tau)
      \nonumber
  \\
     & & -V(x)\psi^{R}(x,\tau).
\end{eqnarray}
We assume uncorrelated finite-width randomness of the potential,
\begin{equation}\label{eq:18.1}
  \overline{V(x)V(x')}=\Delta^{2}\delta(x-x'),
\end{equation}
where the overbar denotes the random average and $\Delta$ is the width 
of the random distribution. 

The $d$-dimensional generalization of Eq.~(\ref{eq:18}) was studied 
in Refs.~\cite{Chen96,Nelson97} via the ``Cole-Hopf transformation,''
\begin{equation}\label{eq:19}
  \psi^{R}(x,\tau) 
  =\exp\left[-\frac{\Phi(x,\tau)}{\hbar}
               +\frac{g^{2}}{2m\hbar}\tau\right].
\end{equation}
The ``Cole-Hopf transformation'' is just another name for the WKB 
method.
The second term in the exponent of Eq.~(\ref{eq:19}) is added 
in order to offset the ground-state energy
(which has no effect in the physics of flux line).
The equation for $\Phi$ is
\begin{equation}\label{eq:20}
  \frac{\partial\Phi}{\partial\tau}=
  -\frac{g}{m}\frac{\partial\Phi}{\partial x}+V(x)
  +\frac{\hbar}{2m}\frac{\partial^{2}\Phi}{\partial x^{2}}
  -\frac{1}{2m}\left(\frac{\partial\Phi}{\partial x}\right)^{2}.
\end{equation}
To see the relevance of each term in the long-distance limit, we 
change the scale as part of a renormalization-group calculation, 
according to
\begin{eqnarray}\label{eq:21}
    x & = & s\tilde{x},
  \\
    \tau & = & s^{z}\tilde{\tau},
  \\
    \Phi & = & s^{\alpha}\tilde{\Phi},
\end{eqnarray}
where the exponents $z$ and $\alpha$ are determined below.
Thus we have
\begin{eqnarray}\label{eq:22}
    \frac{\partial\tilde{\Phi}}{\partial\tilde{\tau}} & = & 
    -s^{z-1}\frac{g}{m}\frac{\partial\tilde{\Phi}}{\partial\tilde{x}}
    +s^{z-\alpha-1/2}\tilde{V}(\tilde{x})
  \nonumber \\
     &  & +s^{z-2}\frac{\hbar}{2m}
     \frac{\partial^{2}\tilde{\Phi}}{\partial\tilde{x}^{2}}
          -s^{z+\alpha-2}\frac{1}{2m}
     \left(\frac{\partial\tilde{\Phi}}{\partial\tilde{x}}\right)^{2}.
\end{eqnarray}
The rescaled random potential is defined by 
$\tilde{V}(\tilde{x})\equiv s^{1/2}V(s\tilde{x})$ so that it 
satisfies 
$\overline{\tilde{V}(\tilde{x})\tilde{V}(\tilde{x}')}
=\Delta^{2}\delta(\tilde{x}-\tilde{x}')$.

The first term of the right-hand side of Eq.~(\ref{eq:22}) 
is a drift term and the second term is the random potential term.
To keep these two terms fixed in the long-distance limit $s\to\infty$, we 
set $z=1$ and $\alpha=1/2$.
The third and the fourth terms are then irrelevant variables in a 
perturbative renormalization group like that constructed in 
Ref.~\cite{Nelson97}.
Thus a Gaussian fixed point controls the physics of what later
turns out to be the regime $g>g_{c2}$.

Upon defining renormalized parameters by
\begin{eqnarray}\label{eq:23}
    \tilde{m} & = & s^{-z-\alpha+2} m,
  \\
    \tilde{\hbar} & = & s^{-\alpha}\hbar,
  \\
    \tilde{g} & = & s^{1-\alpha}g,
  \\
    \mbox{and}\qquad
    \tilde{\Delta} & = & s^{z-\alpha-1/2}\Delta,
\end{eqnarray}
we arrive at a Langevin-type equation in the long-distance limit,
\begin{equation}\label{eq:24}
  \left(\frac{\partial}{\partial\tau}
  +\frac{g}{m}\frac{\partial}{\partial x}\right)\Phi(x,\tau)=V(x),
\end{equation}
where we have dropped the ``tilde'' symbol from all quantities.
Since the ground-state energy was already offset in Eq.~(\ref{eq:19}),
we can eliminate the time derivative by moving onto a new set of 
coordinate as $(x,\tau)\longrightarrow(x,\tau-(m/g)x)$.
We thus see that the solution is a random walk evolving 
into the $x$ direction:
\begin{equation}\label{eq:25}
  \Phi(x,\tau)\equiv\Phi(x)=\frac{m}{g}\int^{x}V(x')dx'.
\end{equation}
The stationary right-eigenfunction in the long-distance limit
is hence given by
\begin{equation}\label{eq:26}
  \phi_{\rm gs}^{R}(x)
  =\exp\left[-\frac{m}{g\hbar}\int^{x}V(x')dx'\right]
\end{equation}
except a normalization factor.
This is equivalent to the ground-state ($k=0$) solution of 
Silvestrov's 
calculations~\cite{Silvestrov98,Silvestrov98-2,Silvestrov98b}.
Equations~(\ref{eq:6}) and~(\ref{eq:26}) then 
give the left-eigenfunction as
\begin{equation}\label{eq:27}
  \phi_{\rm gs}^{L}(x)
  =\exp\left[\frac{m}{g\hbar}\int^{x}V(x')dx'\right].
\end{equation}

Note that the random-walk behavior disappears for the product
$\phi_{\rm gs}^{L}\phi_{\rm gs}^{R}$,
which is completely homogeneous in this approximation.
This is consistent with 
the numerical result in Fig.~\ref{fig:wf-gs} for $g>g_{c2}$.

\subsection{Sample-to-sample fluctuations}

Silvestrov~\cite{Silvestrov98,Silvestrov98-2,Silvestrov98b} 
referred to the above
behavior of $\phi_{\rm gs}^{R}$ and $\phi_{\rm gs}^{L}$
as ``stochastic localization'' (or simply as 
``localization'' in some sentences).
Although the behavior is quite different than the smooth delocalized 
behavior of their product, it is not clear to us whether such states 
should be called ``localized'' either.

\begin{figure}
\epsfxsize=3.375in
\epsfbox{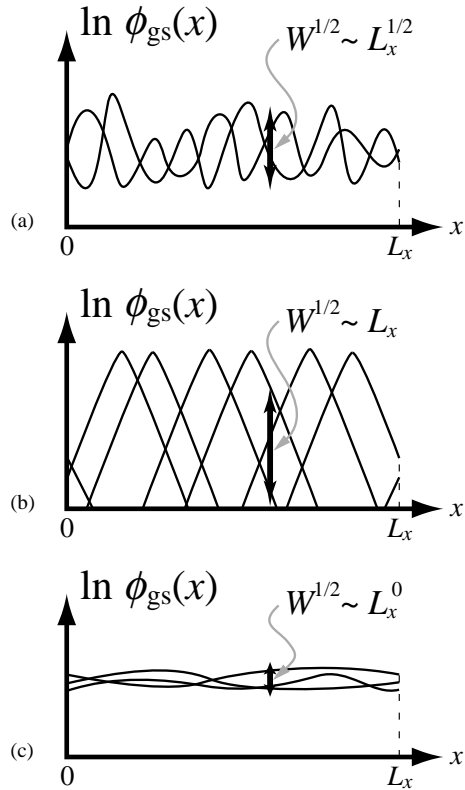}
\caption{Schematic views of sample-to-sample fluctuation $W(L_{x})$ 
for various types of ground state: 
(a) ``random-walk'' wave functions of the form~(\protect\ref{eq:26});
(b) a set of wave functions localized in a conventional sense;
(c) wave functions which are extended in a conventional sense.
Different curves in each graph indicate wave functions of different samples.}
\label{fig:fluc}
\end{figure}
\vskip 0.3in

To stress this point further, we follow Refs.~\cite{Chen96,Nelson97}
and consider the sample-to-sample fluctuations of the logarithm of 
the ground-state wave function at a fixed location, 
\begin{equation}\label{eq:29-5}
  W(L_{x}) \equiv \overline{\ln\phi_{\rm gs}(x)^{2}}
  -\overline{\ln\phi_{\rm gs}(x)}^{2},
\end{equation}
where $L_{x}$ is the system size in the $x$-direction and
$\phi_{\rm gs}(x)$ is either the right- or the left-eigenfunction
of the ground state.
(We choose a normalization such that 
$\int\phi_{\rm gs}^{L}\phi_{\rm gs}^{R} dx=1$.)
The $x$-dependence of this quantity should disappear owing to the 
statistical translational invariance.
As is shown below and in Fig.~\ref{fig:fluc}, we would have
\begin{itemize}
    \item  $W(L_{x})\simeq O(L_{x})$
      for the ground-state wave functions~(\ref{eq:26}) 
      and~(\ref{eq:27});

    \item  $W(L_{x})\simeq O(L_{x}^{2})$
      for a conventional (Hermitian) localized ground state;

    \item  $W(L_{x})\simeq O(L_{x}^{0})$
      for a conventional extended ground state.
\end{itemize}
From this point of view, 
the random-walk behavior of $\phi^{R}$ and $\phi^{L}$ may be viewed
``intermediate'' between localized and delocalized.

The first $L_{x}$-dependence of the quantity $W(L_{x})$ 
is derived either from the wave function~(\ref{eq:26}) or~(\ref{eq:27}) 
as
\begin{equation}\label{eq:30}
    W(L_{x}) \propto\int\!\!\!\!\int dx'  dx'' \,\overline{V(x')V(x'')}
   =  O(L_{x}).
\end{equation}
Next, to calculate $W(L_{x})$ for a conventional localized ground state, 
we assume its asymptotic form as $\phi(x)\sim\exp(-\kappa|x-x_{c}|)$ and 
that the value of $\kappa$ is approximately equal for all samples 
but the localization center $x_{c}$ is different from sample to sample;
see Fig.~\ref{fig:fluc}~(b).
The random average in Eq.~(\ref{eq:29-5}) then reduces to the 
average over $x_{c}$.
Thus we have
\begin{eqnarray}\label{eq:32}
    W(L_{x}) & \sim & \frac{\kappa^{2}}{L_{x}}
    \int|x-x_{c}|^{2}dx_{c}
    -\left[\frac{\kappa}{L_{x}}\int|x-x_{c}|dx_{c}\right]^{2}
      \nonumber
  \\
     & = & O(L_{x}^{2}).
\end{eqnarray}
Finally, the logarithm of a conventional extended ground state should be 
approximately homogeneous in space for all samples and hence have little 
sample-to-sample fluctuation as illustrated in Fig.~\ref{fig:fluc}~(c).
This is the behavior of $\phi_{\rm gs}^{L}\phi_{\rm gs}^{R}$ when 
$g>g_{c2}$.

\section{summary}
\label{sec5}

In conclusion, we have argued that, contrary to some recent statements 
in the literature, delocalization does appear in the eigenfunctions of 
Hamiltonians such as~(\ref{eq:ham1D}) and~(\ref{eq:ham-mat}) when the 
spectrum becomes complex, provided one studies the product of left- 
and right-eigenfunctions.
Delocalization defined by this criterion is consistent with earlier 
definitions based on ability of states with complex eigenvalues to 
carry a nonzero current and the sensitivity to boundary 
conditions~\cite{Hatano96,Hatano97}.

The left- and right-eigenfunctions considered separately provide 
interesting information about the physics of an isolated vortex line 
at its entry and exit points.
Similar conclusions apply to interacting arrays of vortices.
See Sec.~VIII of Ref.~\cite{Hatano97} for a discussion of this 
non-Hermitian many-body problem.
An interesting investigation of tilted interacting vortices at the 
entry and exit boundaries has been initiated by 
Silvestrov~\cite{Silvestrov98b}.

\section*{Acknowledgments}

It is a pleasure to acknowledge helpful conversations with N.~Shnerb, 
A.~Zee and B.I.~Halperin.
This research was supported by the National Science Foundation through 
Grant No.~DMR97-14725 and by the Harvard Materials Research Science 
and Engineering Laboratory through Grant No.~DMR94-00396.




\end{multicols}

\begin{thebibliography}{99}

\bibitem{Hatano96}
N. Hatano and D.R. Nelson,
Phys. Rev. Lett. {\bf 77}, 570 (1996).

\bibitem{Hatano97}
N. Hatano and D.R. Nelson,
Phys. Rev. B {\bf 56}, 8651 (1997).

\bibitem{Chen96}
L.-W. Chen, L. Balents, M.P.A. Fisher and M.C. Marchetti,
Phys. Rev. B {\bf 54}, 12798 (1996).

\bibitem{Shnerb97}
N. Shnerb,
Phys. Rev. B {\bf 55}, R3382 (1997).

\bibitem{Efetov97}
K.B. Efetov,
Phys. Rev. Lett. {\bf 79}, 491 (1997);
Phys. Rev. B {\bf 56}, 9630 (1997).

\bibitem{Feinberg97a}
J. Feinberg and A. Zee,
Nucl. Phys. B {\bf 504} [FS], 579 (1997).

\bibitem{Janik97c}
R.A. Janik, M.A. Nowak, G. Papp and I. Zahed,
Report No. cond-mat/9705098.

\bibitem{Brouwer97}
P.W. Brouwer, P.G. Silvestov and C.W.J. Beenakker,
Phys. Rev. B {\bf 56}, R4333 (1997).

\bibitem{Feinberg97c}
J. Feinberg and A. Zee,
Phys. Rev. E, in press (Report No. cond-mat/9706218).

\bibitem{Goldsheid97}
I.Ya. Goldsheid and B.A. Khoruzhenko,
Phys. Rev. Lett. {\bf 80}, 2897 (1998).

\bibitem{Brezin97}
E. Brezin and A. Zee,
Nucl. Phys. B {\bf 509} [FS], 599 (1998).

\bibitem{Nelson97}
D.R. Nelson and N. Shnerb,
Phys. Rev. E {\bf 58}, No. 2 (1998).

\bibitem{Feinberg97d}
J. Feinberg and A. Zee,
Report No. cond-mat/9710040.

\bibitem{Janik97e}
R.A. Janik, M.A. Nowak, G. Papp and I. Zahed,
Report No. hep-ph/9710103.

\bibitem{Zee97}
A. Zee,
Physica A, in press (Report No. cond-mat/9711114).

\bibitem{Hatano98}
N. Hatano,
Physica A, in press (Report No. cond-mat/9801283).

\bibitem{Shnerb97b}
N.M. Shnerb,
Phys. Rev. B {\bf 57}, 8571 (1998).

\bibitem{Mudry97}
C. Mudry, B.D. Simons and A. Altland,
Report No. cond-mat/9712103.

\bibitem{Shnerb98}
N.M. Shnerb and D.R. Nelson,
Phys. Rev. Lett. {\bf 80}, No. 23 (1998).


\bibitem{Silvestrov98}
P.G. Silvestrov,
Report No. cond-mat/9802219v1.

\bibitem{Silvestrov98-2}
P.G. Silvestrov,
Report No. cond-mat/9802219v2.

\bibitem{Jiang94}
W. Jiang, N.-C. Yeh, D.S. Reed, U. Kriplani, D.A. Beam, M. 
Konczykowski, T.A. Tombrello and F. Holtzberg,
Phys. Rev. Lett. {\bf 72}, 550 (1994).

\bibitem{Safar96}
H. Safar, S.R. Foltyn, Q.X. Jia and M.P. Maley,
Phil. Mag. B {\bf 74}, 647 (1996).

\bibitem{Obaidat97}
I.M. Obaidat, S.J. Park, H. Safar and J.S. Kouvel,
Phys. Rev. B {\bf 56}, R5774 (1997).

\bibitem{Nelson93}
D.R. Nelson and V. Vinokur,
Phys. Rev. B {\bf 48}, 13060 (1993).

\bibitem{Silvestrov98b}
P.G. Silvestrov,
Report No. cond-mat/9804093.


\end{thebibliography}
\end{document}